\documentclass[twocolumn,longauth]{aa} 
\usepackage{graphicx}
\usepackage{txfonts}
\usepackage{natbib}
\bibpunct{(}{)}{;}{a}{}{,} 
\begin{document}
\newcommand{\masq}{magn \arcsec$^{-2}$}
\newcommand{\Jykms}{Jy\,km\,s$^{-1}$}
\newcommand{\mjyb}{mJy beam$^{-1}$}
\newcommand{\HO}{$H_0$}
\newcommand{\kms}{km\,s$^{-1}$}
\newcommand{\acm}{cm$^{-2}$}
\newcommand{\teff}{$T_{eff}$} 
\newcommand{\kmsmpc}{km\,s$^{-1}$\,Mpc$^{-1}$}
\newcommand{\mjb}{mJy beam$^{-1}$}
\newcommand{\jb}{Jy beam$^{-1}$}
\newcommand{\mjbc}{mJy beam$^{-1}$ channel$^{-1}$}
\newcommand{\jykms}{Jy km s$^{-1}$}
\newcommand{\msun}{{M$_\odot$}}
\newcommand{\msunyear}{{M$_\odot$\,yr$^{-1}$}}
\newcommand{\lsun}{{$L_{\odot,B$}}}
\newcommand{\lb}{{L$_{\rm B}$}}
\newcommand{\mlsun}{{\rm M}$_\odot/{\rm L}_{\rm B_{\odot}$}}
\newcommand{\kmsMpc}{km s$^{-1}$ Mpc$^{-1}$}
\newcommand{\hi}{H{\small I}}
\newcommand{\MHI}{M$_{\rm HI}$}
\newcommand{\m}{\hbox{$^{\rm m}$}}
\newcommand{\s}{\hbox{$^{\rm s}$}}
\newcommand{\h}{\hbox{$^{\rm h}$}}
\newcommand{\dn}{D$_{\rm n}$4000}
\newcommand{\oii}{[O{\small II}]}
\newcommand{\oiii}{[O{\small III}]}
\newcommand{\neiii}{[Ne{\small III}]} 
\newcommand{\ie}{{i.e.\,}}
\newcommand{\eg}{{e.g.\,}}

   \title{The VIMOS VLT Deep Survey: \thanks{Based on data obtained
   with the European Southern Observatory Very Large Telescope,
   Paranal, Chile, program 070.A-9007(A), and on data obtained at the
   Canada-France-Hawaii Telescope, operated by the CNRS in France,
   CNRC in Canada and the University of Hawaii.}}

   \subtitle{Tracing the galaxy stellar mass assembly history over the last
   8~Gyr}
   
   \titlerunning{Tracing the galaxy stellar mass assembly history over the
   last 8~Gyr}

\author{
     D. Vergani  \inst{1}\thanks{\emph{Present address:} Universit\`a di Bologna, Dipartimento
   di Astronomia, Via Ranzani 1, I-40127, Bologna, Italy}
\and M. Scodeggio \inst{1}
\and L. Pozzetti    \inst{2} 
\and A. Iovino \inst{3}
\and P. Franzetti \inst{1}
\and B. Garilli \inst{1}
\and G. Zamorani \inst{2} 
\and D. Maccagni \inst{1}
\and F. Lamareille \inst{2}
\and O. Le F\`evre \inst{4}
\and S. Charlot \inst{5,6}
\and T. Contini \inst{7}
\and L. Guzzo \inst{3}
\and D. Bottini \inst{1}
\and V. Le Brun \inst{4}
\and J.P. Picat \inst{7}
\and R. Scaramella \inst{8,9}
\and L. Tresse \inst{4}
\and G. Vettolani \inst{8}
\and A. Zanichelli \inst{8}
\and C. Adami \inst{4}
\and S. Arnouts \inst{4}
\and S. Bardelli  \inst{2}
\and M. Bolzonella  \inst{2} 
\and A. Cappi    \inst{2}
\and P. Ciliegi    \inst{2}  
\and S. Foucaud \inst{10}
\and I. Gavignaud \inst{11}
\and O. Ilbert \inst{12}
\and H.J. McCracken \inst{6,13}
\and B. Marano     \inst{14}  
\and C. Marinoni \inst{15}
\and A. Mazure \inst{4}
\and B. Meneux \inst{1,2}
\and R. Merighi   \inst{2} 
\and S. Paltani \inst{16,17}
\and R. Pell\`o \inst{7}
\and A. Pollo \inst{4,18}
\and M. Radovich \inst{19}
\and E. Zucca    \inst{2}
\and M. Bondi \inst{8}
\and A. Bongiorno \inst{14}
\and J. Brinchmann \inst{20}
\and O. Cucciati \inst{3,21}
\and S. de la Torre \inst{4}
\and L. Gregorini \inst{8,14}
\and E. Perez-Montero \inst{7}
\and Y. Mellier \inst{6,13}
\and P. Merluzzi \inst{19}
\and S. Temporin \inst{22}
}

   \offprints{\mbox{D. Vergani}, \email{daniela.vergani at oabo.inaf.it}}

 \institute{IASF-INAF, via Bassini 15, I-20133, Milano, Italy
   \email{daniela.vergani@oabo.inaf.it}
\and 
INAF-Osservatorio Astronomico di Bologna, Via Ranzani 1, I-40127, Bologna,
Italy 
\and 
INAF-Osservatorio Astronomico di Brera, Via Brera 28, I-20021, Milan, Italy 
\and
Laboratoire d'Astrophysique de Marseille, UMR 6110 CNRS-Universit\'e de
Provence, BP8, F-13376 Marseille Cedex 12, France \and 
Max-Planck-Institut f\"ur Astrophysik, D-85741, Garching, Germany \and 
Institut d'Astrophysique de Paris, UMR 7095, 98 bis Bvd Arago, F-75014 Paris,
France \and 
Laboratoire d'Astrophysique de Toulouse/Tabres (UMR5572), CNRS, Universit\'e
Paul Sabatier - Toulouse III, Observatoire Midi-Pyr\'en\'ees, 14 av. E. Belin,
F-31400, Toulouse, France \and 
IRA-INAF, Via Gobetti, 101, I-40129, Bologna, Italy \and 
INAF-Osservatorio Astronomico di Roma, Via di Frascati 33, I-00040, Monte
Porzio Catone, Italy \and 
School of Physics \& Astronomy, University of Nottingham, University Park,
Nottingham, NG72RD, UK \and 
Astrophysical Institute Potsdam, An der Sternwarte 16, D-14482, Potsdam,
Germany \and 
Institute for Astronomy, 2680 Woodlawn Dr., University of Hawaii, Honolulu,
Hawaii, 96822, USA \and
Observatoire de Paris, LERMA, 61 Avenue de l'Observatoire, F-75014, Paris,
France \and 
Universit\`a di Bologna, Dipartimento di Astronomia, Via Ranzani 1, I-40127,
Bologna, Italy 
\and Centre de Physique Th\'eorique, UMR 6207 CNRS-Universit\'e de Provence,
F-13288, Marseille, France \and 
Integral Science Data Centre, ch. d'\'Ecogia 16, CH-1290, Versoix, Switzerland
\and
Geneva Observatory, ch. des Maillettes 51, CH-1290, Sauverny, Switzerland \and
Astronomical Observatory of the Jagiellonian University, ul Orla 171, PL-30-244
Krak{\'o}w, Poland \and 
INAF-Osservatorio Astronomico di Capodimonte, Via Moiariello 16, I-80131,
Napoli, Italy \and 
Centro de Astrof{\'{i}}sica da Universidade do Porto, Rua das Estrelas,
P-4150-762 Porto, Portugal \and
Universit\`a di Milano-Bicocca, Dipartimento di Fisica, Piazza delle Scienze
3, I-20126 Milano, Italy \and
Laboratoire AIM, CEA/DSM - CNRS - Universit\'e\ Paris Diderot, IRFU/SAp,
F-91191 Gif sur Yvette, France}

   \date{Received ---; accepted ---}

  \abstract {} {Our aim is to investigate the history of mass assembly for
galaxies of different stellar masses and types.}  {We selected a mass-limited
sample of 4048 objects from the VIMOS VLT Deep Survey (VVDS) in the redshift
interval $0.5 \le z \le 1.3$. We then used an empirical criterion, based on the
amplitude of the 4000~\AA\ Balmer break (\dn), to separate the galaxy
population into spectroscopically early- and late-type systems. The equivalent
width of the \oii3727 line is used as proxy for the star formation
activity. We also derived a type-dependent stellar mass function in three
redshift bins.}
{We discuss to what extent stellar mass drives galaxy evolution, showing for
the first time the interplay between stellar ages and stellar masses over the
past 8~Gyr. Low-mass galaxies have small \dn\ and at increasing stellar mass,
the galaxy distribution moves to higher \dn\ values as observed in the local
Universe. As cosmic time goes by, we witness an increasing abundance of
massive spectroscopically early-type systems at the expense of the late-type
systems.
This spectral transformation of late-type systems into old massive galaxies at
lower redshift is a process started at early epochs ({\it z}~$>$~1.3) and
continuing efficiently down to the local Universe. This is also confirmed by
the evolution of our type-dependent stellar mass function.  The underlying
stellar ages of late-type galaxies apparently do not show evolution, most
likely as a result of a continuous and efficient formation of new stars.
All star formation activity indicators consistently point towards a star
formation history peaked in the past for massive galaxies, with little or no
residual star formation taking place in the most recent epochs. In contrast,
most of the low-mass systems show just the opposite characteristics, with
significant star formation present at all epochs.
The activity and efficiency of forming stars are mechanisms that depend on
galaxy stellar mass, and the stellar mass assembly becomes progressively less
efficient in massive systems as time elapses.
The concepts of star formation downsizing and mass assembly downsizing
describe a single scenario that has a top-down evolutionary pattern in how the
star formation is quenched, as well as how the stellar mass is grown.
The role of (dry) merging events seems to be only marginal at {\it z}~$<$~1.3,
as our estimated efficiency in stellar mass assembly can possibly account for
the progressive accumulation of observed passively evolving galaxies.}
{} 
  \authorrunning{D. Vergani et al.}  \titlerunning{Galaxy assembly} \maketitle

\section{Introduction}
\label{sec:intro}

How galaxies form and evolve with cosmic time is one of the key questions in
observational cosmology. A fundamental role towards answering this question is
played by deep surveys sampling over a thousand galaxies on large portions of
the sky.
The global star formation history is reasonably well known out to very high
redshifts and with considerable details up to {\it z}~$\sim$~1 \cite[see for a
recent summary][]{hop06}, but the role of the various physical mechanisms
contributing to the assembly of galaxy stellar mass is still unclear, as is
their importance at different epochs.

In  agreement  with  the  first  formulation  introduced  by  \cite{cow96},  a
downsizing  scenario  for  galaxy  formation persistently emerges  in  several
observational studies \citep{be00, gav02, kod04, bau05, feu05a, feu05b, jun05,
bun06,  bor06,   cim06,  cuc06,  poz07}.    However,  observed  galactic-scale
properties like the  quenching of star formation activity  in massive galaxies
not yet have been reproduced well within the $\Lambda{\rm  CDM}$ framework
\citep[\eg][]{del06,bow06} that successfully describes the hierarchical growth
of dark matter halos and the galaxy clustering properties.
 
The observed discrepancies between models and observations are probably due to
our limited ability to reproduce the actual physics on galactic scales with
simple recipes. Perhaps some feedback mechanisms are still missing in models.
Thus, a phenomenological approach to the problem of the stellar mass assembly
as a function of cosmic time, \ie one that relies as much as possible on
observational data, can suggest new implementations in current semi-analytical
models.

In this paper we use the first epoch observations of the VIMOS-VLT Deep Survey
\citep[VVDS,][]{lef05} to derive stellar masses, stellar ages and star
formation efficiency for a statistically significant sample of galaxies.  We
analyse how stellar mass is assembled in active and quiescent galaxies
classified following a spectroscopic scheme, and how this process evolves with
cosmic time using spectroscopic diagnostics. This analysis provides
constraints on the hierarchical scenario thought to govern dark matter
assembly and can be pursued using VVDS data starting from a time when the
Universe had $\sim$~30\% of its present age.

The present work is organised as follows: the data and sample selection are
presented in Section\,2; the methodology we adopted to perform our analysis is
described in Section\,3. Results are given in Sect.\,4 and a summary in
Sections 5. Throughout this work we assume a standard cosmological model with
$\Omega_M = 0.3$, $\Omega_\Lambda = 0.7$ and $H_0 = 70 \, \mathrm{km} \,
\mathrm{s}^{-1} \, \mathrm{Mpc}^{-1}$.  Magnitudes are given in the AB system.

\section{The first epoch VVDS sample}
\label{sec:data}

The primary observational goal of the VVDS as well as the survey strategy are
presented in detail in \citet{lef05}. Here we stress that in order to minimise
possible selection biases, the VVDS has been conceived as a purely
flux-limited survey, \ie no target pre-selection according to colours or
compactness is implemented.  In this paper we consider the deep spectroscopic
sample of the first epoch data in the VVDS-0226-04 field (from now on simply
VVDS-F02) that targets objects in the magnitude range $17.5\leq{\rm
I_{AB}}\leq24$.

First-epoch spectroscopic observations in the VVDS-F02 field were carried out
with the VIMOS multi-object spectrograph \citep{lef05}.  The LRred grism was
used with 1\arcsec\ wide slits to cover the spectral range
5500~\AA$<\lambda<$~9400~\AA with an effective spectral resolution {\rm
R}~$\sim$~230 at $\lambda$~$=$~7500~\AA.  The spectroscopic targets were
selected from the photometric catalogues using the VLT-VIMOS Mask Preparation
Software \citep[VVMPS;][]{bot05}.  The spectroscopic multi-object exposures
were reduced using the VIPGI tool \citep{sco05, zan05}. Further details on
observations and data reduction are given in \citet[]{lef04b,lef04a}.

The first-epoch data sample in the VVDS-F02 field extends over a sky area of
$0.7 \times 0.7 \deg^2$ and has a median redshift of {\it z}~$\sim$~0.76.  It
contains 7267 objects with secure redshifts, corresponding to an average
random sampling rate of 23\% of the initial photometric sample.
It is important to stress that we explored the VVDS-F02 redshift success rate
for different galaxy types (from star-forming to quiescent galaxies) in
\citet{fra07} obtaining a similar fraction of galaxy types in the photometric
sample and in the spectroscopic sample. Thus we have no significant bias
against special types of galaxies in our survey as our ability to obtain
redshifts for star-forming and quiescent galaxies are similar.

In this work we used the \oii$\lambda$3727 line and the 4000~\AA\ break as
indicators of star formation activity and stellar ages respectively.  As a
consequence, we limited our sample to galaxies within the $0.5 \le z \le 1.3$
redshift interval, giving us a final number of 4277 galaxies, after removing
spectroscopically confirmed broad line AGNs and stars.

This sample provides an excellent laboratory to analyse spectral properties of
galaxies at several redshift ranges, especially as it is accompanied by a
wealth of photometric ancillary data, collected at several telescopes and
described in the following papers: \citet[BVRI bands at CFHT,][]{mcc03};
\citet[U-band at ESO-2.2m/WFI,][]{rad04}; $ubvrz$ bands by the CFHT Legacy
Survey project \citep{mcc07}, and J and K$_s$ at ESO-NTT/SOFI
\citep{iov05,tem06}.

\section{Methodology}
\label{sec:met}

In this section we will describe in some details the methodology adopted to
obtain measurements of \oii$\lambda$3727 and \dn\ from VVDS spectra and to
derive estimates for specific star formation rates (SSFRs) and stellar masses.

\subsection{Spectral measurements: \oii\ and \dn}
\label{sec:spec}

Spectral features measurements have been obtained for all galaxies in our
sample using the {\it platefit$\_$vvds} software package. This software
implements the spectral feature measurement techniques described by
\citet{tre04}, but takes into account the lower spectral resolution of our
data. Here we give a brief outline of the fitting procedure and we refer to
Lamareille et al. (2007) for further details.  As a first step, the best
fitting stellar continuum derived from a grid of stellar population synthesis
models (Bruzual \& Charlot 2003, hereafter BC03) is subtracted from the
observed spectrum.  Any remaining residual is removed by fitting a low-order
polynomial to the continuum-subtracted spectrum, and then all emission lines
are fitted simultaneously with a Gaussian profile. Finally, absorption
features and spectral breaks are measured after having subtracted
emission-lines from the original spectrum.

The two spectral measurements we use in this paper are the equivalent width of
the \oii$\lambda$3727 doublet (EW\oii\ from now onwards) and the amplitude of
the 4000~\AA\ break.  In particular, we adopt the so-called narrow definition
for the 4000~\AA\ break (hereafter \dn), introduced by \citet{bal99} and based
on the ratio of the average spectral flux density in the bands
$4050-4250$~\AA\ and $3750-3950$~\AA\ around the break.  Compared to the
original definition introduced by \citet{bru83} the newly proposed index has
the advantage of reducing the effect of dust reddening, because of the
narrower portion of galaxy spectrum involved in the measurement \citep[see for
details][]{kau03a}.

Comparing repeated observations available for 140 objects in our sample, we
estimated that the typical relative uncertainty affecting our measurements is
$\sim$~27\% and $\sim$~10\% for the EW\oii\ and \dn, respectively.  Given the
typical signal-to-noise ratio of the spectra, we estimate an average detection
threshold for the \oii\ doublet of 8~\AA.
Since the \dn\ accuracy can depend on galaxy redshift because some fringing
  residuals contribute to increase the noise in the VVDS spectra, we compared
  the repeated \dn\ measurements separately for galaxies with {\it z}~$<$~0.9
  (where the measurement is not affected by fringing) and for galaxies with
  {\it z}~$>$~0.9 (where our measurements begin to be affected by fringing).
  The scatter of the \dn\ values in the two redshift bins is always within the
  quoted values. Therefore we can safely conclude that fringing does not
  significantly affect our results.

 \subsection{Spectral classification}
 \label{sec:class}

To study the role played by the stellar mass in regulating the active phase of
star formation, we adopt a parametric classification of galaxies based on the
\dn\ index used as an estimator of stellar ages \citep{ham85,bal99}.

We choose a value of \dn~$=$~1.5 to distinguish between {\it spectroscopic
late-type} (\dn~$<$~1.5) and {\it spectroscopic early-type} (\dn~$>$1.5)
galaxies.

Our choice is motivated both by galaxy evolution models and
observations. Single stellar population (SSP) models have shown that in a
galaxy with solar metallicity experiencing an instantaneous burst of star
formation, the spectrum of the underlying stellar population reaches values of
\dn\ larger than 1.5 in $\sim$~1~Gyr \citep{kau03a}.  Different results
obtained using different stellar libraries are within the error of the \dn\
measure \citep{kau03a,leb06}, while there is a possible dependency on
metallicity for old stellar ages ($>$~1~Gyr). Galaxies with values of
metallicity as extreme as $0.05$ ($2.5$) ${\rm Z}_{\sun}$ reach values of \dn\
larger than 1.5 in 2.5 (0.8) ~Gyr after a burst \citep{kau03a}.  \cite{leb06}
investigate the \dn\ evolution using other modes of star formation history.
They show that the \dn\ index reaches values larger than 1.5 after 1~Gyr from
the initial activity also in galaxies with a period of constant star formation
truncated 0.5~Gyr after their formation, while galaxies with recurring burst
modes always keep the \dn\ values below 1.5 in a period as long as 9~Gyr.\\
\noindent Observationally in the present-day galaxies, values of \dn\ smaller (larger)
than 1.5 have identified young (old) stellar populations with a separation
between the two populations at 1~Gyr \citep{kau03b}. This limit has been
widely used in other observational studies \citep{mo02,mig05}.

We conclude that the \dn, despite the uncertainties discussed above, is still
completely adequate to provide a broad separation of the galaxy population
into active and quiescent galaxies, although it might not provide a very
precise age estimate for the stellar population.

Hereafter we refer to our {\it spectroscopic early-} and {\it spectroscopic
late-type} galaxies simply as early- and late-type galaxies, although the
classification of a galaxy on colour, morphological, spectroscopic based
schemes should be kept distinct.  The different terminologies should not be
mixed up, despite the robust trends existing among morphologies, colours, gas
content, star formation activity, and other properties \citep[for a recent
review see][]{ren06}.  The category of early-type galaxies, by virtue of their
relatively simple star formation history, are generally used as the
preferential category of objects to test galaxy formation and evolution and
therefore one should be careful in the definition chosen to select them
\citep[see][]{fra07}.

\subsection{Star Formation Rate Estimates}\label{sec:ssfr}

Adopting the formula proposed by \citet{guz97} we can translate the EW\oii\ in
star formation rate estimates:

\begin{equation}\label{eq_SFR}
{\rm SFR(M_{\odot}~yr^{-1})} \approx 10^{-11.6-0.4 ({\rm M_B} - {\rm
M_B}_{\odot})} {\rm EW[O{\small II}]} \,\,\,     ,
\end{equation}

where we used the absolute magnitudes ${\rm M_B}$ as estimated in
\citet{ilb05}.  Although the average accuracy of the absolute
spectrophotometric calibration of the VVDS sample is better than $\sim$~10\%
r.m.s. in optimal weather conditions, we favour the use of the EW\oii\ instead
of line fluxes to estimate the SFR. In this way we minimise the dependency of
our estimates on less optimal weather conditions and slit losses \citep[see
for details][]{lef05}.

There are three main caveats for the reliability of the SFR conversion
obtained from EW\oii\ using Eq.\,1 over the whole redshift interval ($0.5 \le
z \le 1.3$): the uncertainty on the fraction of active galactic nuclei (AGN)
present in our sample, the cosmic evolution of metallicity, and the amount of
extinction.

One risk in interpreting line measurements as proxy for star-formation
activity is the contamination arising from the presence of an AGN in the
central region of a galaxy.  In this case, the emission lines in the spectrum
would be a mixture of contributions from both the AGN and star formation.  We
removed from our final sample the spectroscopically confirmed broad-line AGNs.
We estimate a contamination from type II AGN-dominated galaxies on the order
of 7\% over the redshift range explored.  This fraction has been estimated
using the standard line ratio diagnostic tools, \ie
\oiii$\lambda$5007/H$\beta$ vs. \oii$\lambda$3727/H$\beta$ in the redshift
interval $0.5<z<0.8$ and new diagnostics using \oii$\lambda$3727,
\neiii$\lambda$3869 and H$\delta$ in the redshift interval $0.9 < z < 1.3$
(for details we refer to Lamareille et al. in prep. and Perez-Montero et
al. in prep.). However, given the very significant uncertainty that is
affecting the identification of these narrow-line AGNs and their relatively
small number, we have decided not to remove them from the sample.

Another possible limitation we face in converting EW\oii\ into star formation
estimates is related to the metallicity and its cosmic evolution, that could
result into a systematic change of line intensities with redshift even if the
star formation activity were to remain constant.  \cite{mou05} have shown that
a variation of the metallicity by a factor of 4 can produce a variation of a
factor of about 3 in the reddening-corrected \oii$\lambda3727/{\rm H}\alpha$
flux ratio \citep[see Fig.\,12 in][]{mou05}. This relation however has an
opposite behaviour for different metal regimes.
We have the scenario in which the \oii$\lambda3727/{\rm H}\alpha$ flux ratio
increases with metallicity in metal-poor galaxies which are typically low-mass
objects (log(M/\msun)\,$<$\,10, $\sim$~0.2Z$_{\odot}$), and it decreases with
metallicity in metal-rich galaxies which are typically massive galaxies
(log(M/\msun)\,$>$\,{11}, $1.4Z_{\odot}$), see e.g. \cite{gal05}.
As for the direct effect of metallicity evolution on line strength, there is
still considerable debate about the details of the evolution of the
mass-metallicity relation.  \citet{sav05} find very little change in
metallicity for massive, already high-metallicity systems, up to {\it
z}~$\sim$~1.  They find a significant overall increase of metallicity for
low-mass, generally low-metallicity objects.  On the contrary, \citet{lam06}
observe a general decrease of metallicity which is essentially independent
from galaxy mass, while the VVDS data (Lamareille et al. in prep.) show some
indication of a smaller change of metallicity for low-mass systems than for
high-mass ones.  Overall these studies are concordant in finding an evolution
of the metallicity within a factor of 2 up to {\it z}~$\sim$~1. This would
imply an uncertainty of less than a factor of two in our star formation rate
estimates.  As the observed variations in star formation activity are
significantly larger than this uncertainty, and the metallicity estimates we
can obtain are also highly uncertain because of the low spectral resolution
and low S/N of our spectra, we prefer not to correct our \oii\ measurements
for metallicity variations.

Finally, we also have to consider the effects of dust extinction on our SFR
estimates, although they are reduced by the use of equivalent widths instead
of line fluxes.  The observed line and continuum fluxes involved in the
determination of the equivalent width are both affected by the diffuse dust
extinction within the galaxy, and the only differential effect would be due to
localised extinction within HII regions, that should account for only a minor
fraction of the total dust extinction within a galaxy \citep{ken89,cl01}.  To
account for the dust attenuation we corrected the EW\oii\ by the amount of the
extinction provided by our template fitting using PEGASE \citep[][ see next
section]{peg97}.  The correction values obtained with this method vary within
the interval EW\oii$_{\rm corr~}$/~EW\oii~$\in [1-2.7]$.

To the caveats we already mentioned, for sake of completeness we should also
add the possible uncertainties derived from the unknown physical condition of
the ionizing gas and those related to the difference between our sample and
the one originally used to calibrate H$\alpha$/\oii\ in Eq.\,1 \citep{guz97}.

\subsection{Stellar Mass Estimates}\label{sec:mass}

We obtain the stellar masses for our galaxy sample by fitting the photometric
and spectroscopic data with a grid of stellar population synthesis models
generated by the PEGASE2 population synthesis code \citep[Fioc \&
Rocca-Volmerange 1999, astro-ph/9912179;][]{peg97}, and using the GOSSIP
Spectral Energy Distribution tool \citep{fra05}.  The models were produced
assuming a Salpeter initial mass function \citep{sal55} to allow an easy
comparison with previous analyses \citep[\eg][]{be00, fon04}.  The use of
other prescriptions for initial mass functions (IMFs) does not introduce any
significant difference in the redshift evolution of the mass estimate in the
{\it z} interval explored as the ratio between masses derived with e.g. a
Salpeter IMF and the ones derived with a Chabrier IMF is close to a constant
for most star formation histories (see Pozzetti et al. 2007, hereafter P07).
We use a set of delayed exponential star formation histories \citep[see Eq.\,3
in][]{gav02} with galaxy ages, $t$, in the range from 0.1 to 15~Gyr, and star
formation time-scales, $\tau$, between 0.1 and 25~Gyr. Internal dust
extinction is handled self-consistently with the star formation activity by
the PEGASE2 code and no burst component has been added.

Stellar masses obtained with this method were compared with other estimates
based on the fitting of the photometric data only and photometric plus
spectroscopic data with \citet{bc03} models derived using different families
of star formation histories. We find the various estimates to be consistent
with each other in the redshift interval investigated. We estimate a
statistical uncertainty on the mass estimate of approximately 0.2~dex.  In
particular, our methodology is equivalent to the smooth SFHs method described
in P07 even if they adopt an exponentially decreasing star formation history
and \citet{bc03} models, rather than a delayed exponential star formation
history and PEGASE2 population synthesis models as in this work.  The
photometry coverage was improved with respect to P07, and it now includes
UKIDSS NIR (JK bands) photometry \citep{lau07} and SPITZER-IRAC (3.6,4.5,5.8,
and 8~$\mu$m) photometry from the SWIRE survey \citep{lon03}. Even if our
stellar masses are well consistent with the ones by P07, the new photometry
improves the mass determination avoiding the need for any further statistical
correction, as made in P07 because of the lack of NIR photometry in about half
of the sample.

Stellar mass completeness limits have been estimated using the same stellar
population synthesis models described above to predict apparent I-band
magnitudes for galaxies with different star formation histories (and therefore
different mass to light ratios) as a function of redshift. Comparing these
predicted magnitudes with the magnitude limit used to select the survey
spectroscopic sample we estimate that we can observe galaxies above ${\rm
log(M_{\star})}\simeq {9.6}, {9.9}, {10.5}$~\msun\ for $z\in[0.5-0.7],
[0.7-1], [1-1.3]$, respectively.  These stellar mass limits correspond to a
completeness $>$~80\% for the early-type population and more than 95\% for
late-type galaxies.

We refer to P07 for the method adopted to construct the stellar mass
function. Here we briefly describe the most important points. We derive the
mass functions using the classical non-parametric 1/V$_{max}$ formalism
following \cite{sch68} and \cite{fel76}.  We use for the fit the Schechter
function limiting ourselves above the 80\% completeness stellar mass limits
for the two different population examined. In the estimate of the mass
function we take into account also the incompleteness resulting from
non-targeted sources in our spectroscopic observations and from spectroscopic
failures.

\begin{figure*}[ht!]
\begin{center}
\hspace*{-0.7cm} \includegraphics[width=18.5cm,angle=0]{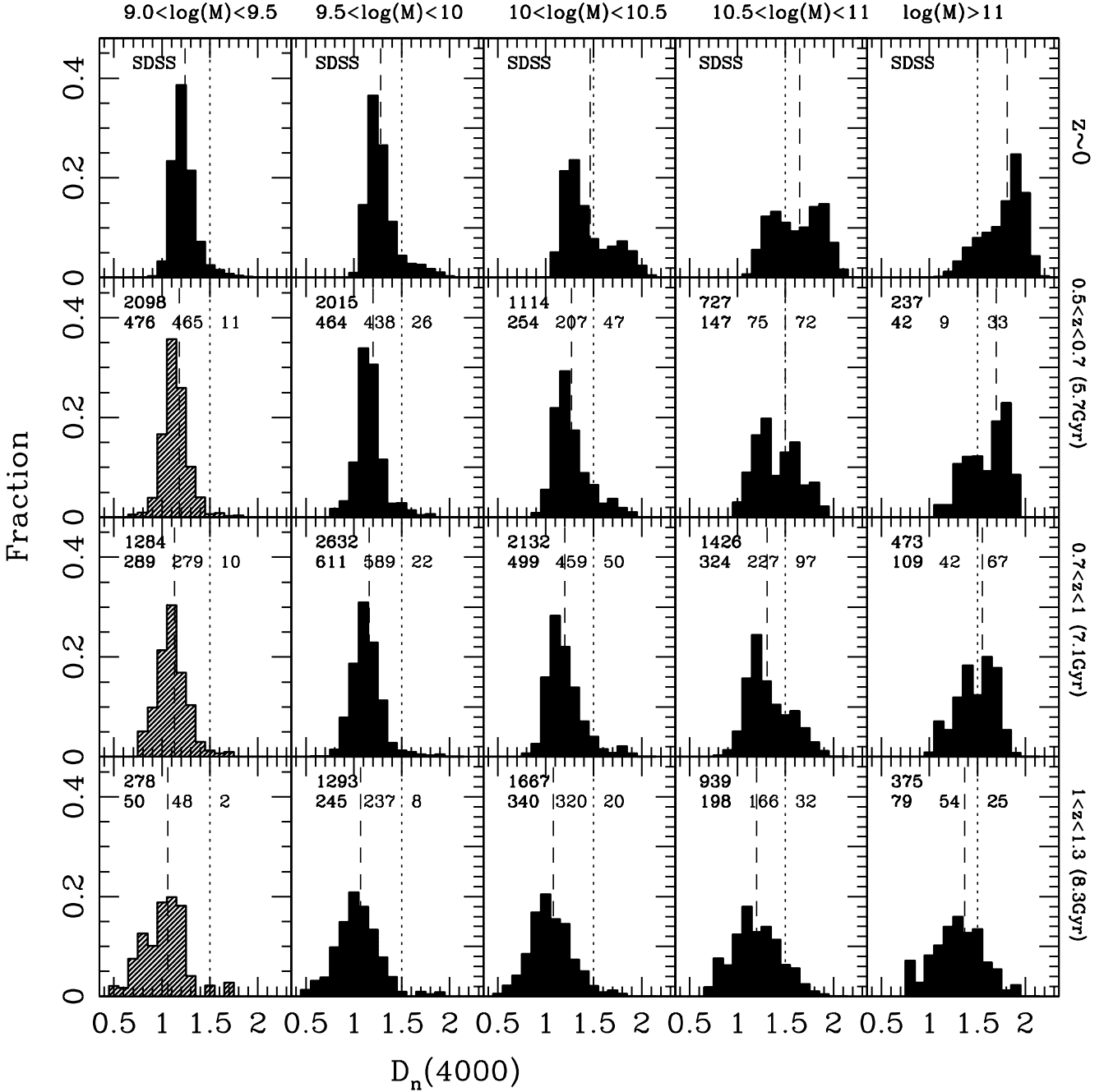}
\end{center}
\vspace*{0.2cm} 
\caption{ Histograms showing the fraction of VVDS-F02 galaxies as a function
of \dn\ in five different ranges of stellar mass and in three redshift bins
($0.5-0.7, 0.7-1.0, 1.0-1.3$, rows 2-4).  We plot as a comparison in the first
row the \dn\ galaxy distribution for the local Universe ({\it z}~$\sim$~0)
from the SDSS~DR4 \citep{DR4}.  The counts corrected for non-targeted sources
in our spectroscopic observations and for spectroscopic failures (unidentified
sources) following \citet{ilb05} are reported in the first line at the top of
each panel, the observed counts of (total, late-, and early-type) galaxies in
the second line. The vertical dotted lines at \dn~$=$~1.5 represent the adopted
separation between early- and late-type galaxies.  The vertical dashed lines
represent the median \dn\ distribution in each panel.  The panels in the first
column dominated mostly entirely by late-type galaxies have dashed \dn\
distribution when the mass completeness for the late-type population is less
than 50\%.}
\label{fig:d4n} 
\end{figure*}

\section{Results}
\label{sec:results}

\subsection{Evolution of Stellar Ages with Stellar Mass}
\label{sec:d4n}

In the local Universe there is a well established observational evidence that
while young galaxies are preferentially low-mass systems, old galaxies are
mostly massive systems.  In particular, \citet{kau03b} identify in their
Fig.\,2 the population of the first peak of the \dn\ distribution observed at
\dn~$\sim$~1.3 as emission line objects with a mean stellar age of 1~--~3 Gyr
and low stellar mass (log(M/\msun)\,$<$\,10.4). The second peak at \dn~$\sim$~1.85
corresponds instead to old ($>10$~Gyr), massive ellipticals.

Figure\,1 shows the histogram of the \dn\ distribution in different intervals
of stellar masses and redshifts as indicated on the top and on the right,
respectively. For the VVDS-F02 data (rows 2-4) the histograms are corrected
for non-targeted sources in our spectroscopic observations and for
spectroscopic failures (the corrected total number and the actual number of
observed galaxies are reported in the first and second lines, respectively, at
the top of each panel).

To allow a direct comparison with the local Universe, the first row of Fig.\,1
shows the \dn\ distribution as derived from a local sample of SDSS~DR4
galaxies \citep{DR4} within the same interval of stellar masses as ours
(properly rescaled to account for the different IMFs).  This plot enables us
to explore the presence of any relation between stellar ages and stellar
masses in the distant Universe.  The vertical dotted lines represent the
cutoff at \dn~=~1.5 adopted to sub-divide our objects in early- and late-types,
as described in the Sect.\,\ref{sec:class}.  The vertical dashed line in each
panel is the median of the \dn\ distribution in that particular stellar mass
interval and redshift bin (see Tab.\,1). In the low-mass interval at high
redshift the incompleteness for early-type galaxies affects the high value
tail of the \dn\ distribution, while it does not strongly affect the median
values, dominated by late-type galaxies.

\begin{table}[ht!]
\begin{center}
\tabcolsep0.4mm 
\caption{\sc {\dn\ median distribution}}
\label{tab:median}
\begin{tabular}{cccccc}
\hline \hline\\ & {$10^{9-9.5}$~\msun} & $10^{9.5-10}$~\msun& $10^{10-10.5}$~\msun&$10^{{10.5-11}}$~\msun&$10^{>{11}}$~\msun\\ 
$z$ & & & & \\ \\[.3ex] \hline
$\sim 0$  &$1.24 \pm0.00$ & $1.28 \pm0.00$ & $1.46 \pm0.00$ & $1.66 \pm0.00$ &$1.81\pm0.00$\\ 
$0.5-0.7$ &$1.18 \pm0.01$ & $1.20 \pm0.01$ & $1.27 \pm0.01$ & $1.50 \pm0.02$ &$1.69\pm0.02$\\ 
$0.7-1.0$ &$1.13 \pm0.01$ & $1.16 \pm0.01$ & $1.21 \pm0.01$ & $1.31 \pm0.01$ &$1.55\pm0.02$\\ 
$1.0-1.3$ &$1.07 \pm0.04$ & $1.07 \pm0.02$ & $1.09 \pm0.01$ & $1.20 \pm0.02$
&$1.37\pm0.02$\\   
\hline \hline\\
\end{tabular}
\begin{minipage}{9cm}
\normalsize{ {\small NOTES:} \dn\ median values in five intervals of stellar
mass and three redshift bins for our VVDS-F02 sample.  The error quoted in
each interval represents the error on the median distribution using a
statistical Jackknife technique.  The median values of the \dn\ distribution
in a nearby galaxy sample (SDSS DR4) are also listed.}
\end{minipage}
\end{center}
\end{table}

The first evidence shown by Fig.\,\ref{fig:d4n} is the different behaviour of
the \dn\ distribution for the five bins of stellar mass explored. They show a
clear dependency of the stellar age as inferred by \dn\ values on the stellar
mass that holds at all epochs up to {\it z}~$\sim$~1.3.  Even at the highest
redshift bin investigated the \dn\ distributions closely follow the behaviour
observed in the local Universe. Low-mass galaxies have small \dn\ values and,
as mass increases, the galaxy distribution moves to larger \dn\ values.  At
any fixed redshift bin, the relative abundance of early-type galaxies
(\dn~$>$~1.5) increases as mass increases.  The largest \dn\ values, and thus
the oldest underlying stellar populations or early-type galaxies following
Sect.\,\ref{sec:class}, are hosted in the most massive galaxies at all
redshifts up to {\it z}~$\sim$~1.3.

Another striking evidence coming from Fig.\,\ref{fig:d4n} is that for the
lower mass galaxies (log(M/\msun)\,$<$\,{10}) the shape of the \dn\
distribution does not seem to evolve with time between {\it z}~$\sim$~1 and
{\it z}~$\sim$~0. For these galaxies the \dn\ distribution is peaked around
\dn~$\sim$~1.1 in all redshift bins, suggesting a good correspondence up to
high redshifts between a low stellar mass and a young underlying stellar
population.
Intermediate and high-mass galaxies show significant evolution with redshift
at {\it z}~$>$~0.7 as testified by the progressive shift of the median value
(dashed line in Fig.\,1).
As cosmic time goes by, for stellar mass larger than log(M/\msun)\,$>$\,10
galaxies populate progressively the locus of larger \dn\ values and a
secondary peak emerges in the \dn\ distribution.  The gradual accumulation of
the secondary peak is a process started at early epochs that continues
efficiently down to the local Universe.

We find an age-stellar mass relation with low-mass galaxies having younger
stellar populations and more massive galaxies possessing older stellar
population. This correlation, originally observed in the local Universe by
e.g. \citet{kau03b}, seems to hold up to {\it z}~$=$~1.3. Our findings on the
progressive accumulation with time of massive galaxies with large \dn\ values
can be interpreted in the framework of a top-down picture of the stellar mass
assembly history of galaxies, also termed {\it assembly downsizing}. With this
terminology we refer to the galactic property that regardless of type, massive
galaxies are fully assembled at earlier times, with less massive galaxies
assembled later (when mass is defined as stellar mass).  Note that here we
make no statement about how the stellar mass was aggregated, as both merging
and star formation can contribute to the assembly process. However we will
argue in Sect.\,\ref{sec:efficency} that galaxy major merger does not seem to
be a fundamental mechanism, at least at {\it z}~$<$~1.
 
Evidence of this top-down stellar mass assembly is already present in
literature, e.g. \citet{tho05, cim06, bun06}.  Studies on stellar age,
metallicity, and cosmic star formation at {\it z}~$\sim$~0 prove that stars in
massive galaxies were formed long ago and over a short period of time
\citep[\eg ][]{tho05,hea04,jim05}.  At high redshifts the reddest galaxies
already possess old ages ($>1$~Gyr), establishing their formation epoch at
{\it z}~$>$~2 \citep{mcc04,cim04}. Furthermore, massive proto-disk galaxies
already exist at {\it z}~$\sim$~2-3 \citep[\eg][]{lab03, sto04, fos06,
gen06}.  On the contrary, the underlying stellar ages of late-type galaxies
apparently do not show evolution.  This observation is probably justified by a
continuous and efficient formation rate of new stars (see next
Sect.\,\ref{sec:mf}) as discussed also in other studies \citep{arn07,fab07}.

Theoretical models, however, predict that stars in an object can be born in a
  different place from where they finally end up (e.g. de Lucia et
  al. 2006). Even if we observe a downsizing signal, stars born long ago in
  several objects could still coalesce into a single object in a hierarchical
  way making the concepts of assembly of a system and assembly of its stars
  different.  However, to be able to test this theoretical prediction we need
  to wait for the next generation of surveys and more detailed semi-analytical
  models.

Based on our phenomenological approach we can conclude as follows.
Our sample shows to which extent the stellar mass drives the galaxy
evolution.  For the first time we present a direct comparison of
stellar ages and stellar masses of galaxies over the last 8~Gyr
extending to higher redshift the seminal work by \citet{kau03b} on
{\it z}~$=$~0 galaxies.

\subsection{SFR dependence on stellar mass and redshift}
\label{sec:o2dep}

A cosmic decline of the star formation activity has been presented in several
studies, starting from field redshift surveys by \citet{lil96,mad96}.
Nowadays, the knowledge of the cosmic star formation history out to redshift
{\it z}~$\sim$~1 is quite profound \citep[][ and references
therein]{hop06}. However many aspects of galaxy evolution, like the
contribution to galaxy mass assembly of various categories of galaxies, still
need to be understood \citep[see for details e.g. ][]{cim06}.

\begin{figure}
\begin{center}
\leavevmode
\includegraphics[width=8.9cm,angle=0]{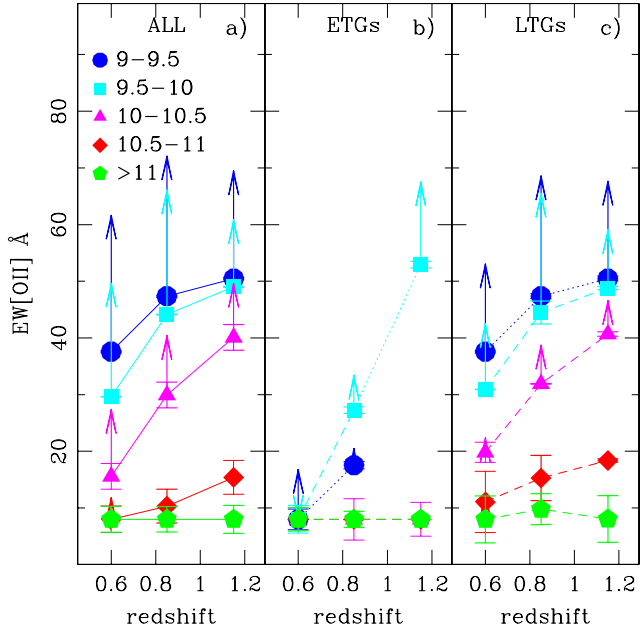}
\end{center}
\caption{The median EW\oii\ as a function of the redshift for our VVDS-F02
  sample ({\bf a}) in bins of stellar masses, coded as follows: $9.0 < {\rm
  log(M/M_{\odot})} \le 9.5$ (blue, circle), $9.5 < {\rm log(M/M_{\odot})} \le
  {10}$ (cyan, square), ${10} < {\rm log(M/M_{\odot})} \le {10.5}$ (magenta,
  triangle), $10.5 < {\rm log(M/M_{\odot})} \le {11}$ (red, diamond), and
  ${\rm log(M/M_{\odot})} >{11}$ (green, pentagon).  The same visualization is
  presented for the spectroscopic early- ({\bf b}) and late-type galaxies
  ({\bf c}). The symbols represent the observed EW\oii\ and the arrows the
  dust corrected EW\oii. The points are connected with dashed (dotted) lines
  when the mass completeness of each population is above (below) 50\%.}
\label{fig:count2}
\end{figure}

Panel a of Fig.\,\ref{fig:count2} shows the median distribution of EW\oii\ in
the redshift interval {\it z}~$=$~0.5-1.3 as a function of the stellar mass. Galaxies
with no EW\oii\ detection are taken into account in the median computation
assigning to them the EW\oii\ upper limit estimated from the 1~$\sigma$
measurement uncertainty. Interpreting the EW\oii\ as a proxy for star
formation activity, we can infer that the latter has a clear redshift
evolution which is strongly modulated by stellar mass. We observe the already
well established global star formation decline from {\it z}~$\sim$~1 to the
present epoch, and also that within a given redshift bin the smaller the
stellar mass of the galaxy, the more actively it is forming stars.

These results are in excellent agreement with the findings by \citet{feu05a}
based on a large sample of 9000 galaxies although with only photometric
redshifts from the FORS Deep Field and the GOODS-S field.  They show a similar
trend for the SSFRs at different stellar mass intervals with no contribution
to the stellar mass growth for massive galaxies \citep[cf. Fig.\,2
in][]{feu05a}. We also find an excellent agreement in the common redshift
range with the study of 207 galaxies from the spectroscopic GDDS
\citep{jun05}. They show that massive GDDS galaxies
(log(M/\msun)\,$>$\,{10.2}) are in a quiescent mode, and lower mass GGDS
galaxies in a burst phase. Similar results have been obtained by \citet{zhe07}
using $\sim$~15,000 COMBO-17 galaxies with photometric redshifts, supplemented
by ultraviolet and infrared data to account for otherwise undetected star
formation activity.

Our results, being obtained on a large spectroscopic sample, put on firmer
grounds the results previously found using photometric galaxy catalogues or
smaller spectroscopic surveys: the star formation in massive systems is at
very low values since {\it z}~$\sim$~1-1.3, while star formation increases
when considering less massive systems at each redshift bin explored.

Panels b and c of Fig.\,\ref{fig:count2} show what happens when one considers
early- and late-type galaxies separately.  
Very massive galaxies (log(M/\msun)\,$>$\,{11}) irrespective of their spectral
type show negligible star formation activity at each redshift bin probed.
Moving to lower masses, galaxies of different spectral classification show
different behaviour (both in slope and normalization of EW\oii\ values). This
result holds over the entire inspected redshift range.
In panel b at masses lower than log(M/\msun)\,$\sim$\,{10} there is an indication of
the presence of galaxies with an old underlying stellar population (ETGs) and
non-negligible levels of \oii.
However, given a number of considerations, including the larger mass
incompleteness and the small number statistics for this class of objects (see
numbers quoted in Fig.\,1), one should be cautious in putting too much weight
on this result.  Interestingly, with our classification scheme we find a lower
contamination of star forming galaxies among early-type objects than reported
in previous studies (5\% vrs. 20-50\%, see \citet{str01, fra07, cim02,
kri06}).  This could be the result of our classification criterion that, being
based on \dn, allows rejecting e.g. dusty-star forming EROs that possess red
colours but low \dn\ values \citep{mig05}.
Another concern we need to stress is the possible contamination from the type
II AGN-dominated galaxies as they might still be present in our sample.  A
number of studies have recently shown that emission lines in early-type
galaxies at {\it z}~$=$~0 are mainly due to AGN rather than to star formation
activity \citep[e.g.][]{wei07}. Therefore we can not exclude the possibility
that even the 5\% of early-type galaxies with non-negligible star formation
could actually be AGNs, thus casting doubt on the possibility of interpreting
EW\oii\ as star formation indicator for low-mass early-type galaxies in panel
b of Fig.\,2.

Panel c of Fig.\,2 shows clear trends for late-type galaxies both as a
function of redshift and of mass. Galaxies have a stronger EW\oii\ going back
in cosmic time and this trend is steeper for less massive galaxies.  The
trends visible in panel a of Fig.\,2 are therefore driven by the late-type
galaxies.

An important piece of information about the global star formation history of
galaxies becomes apparent when comparing the results summarised in
Figs.\,\ref{fig:d4n} and \ref{fig:count2}. It is known that the \dn\ strength
is the result of the cumulative star formation history of a galaxy. As already
discussed in Sect.\,\ref{sec:class}, \dn\ values below 1.5 are indicative of
significant star formation activity within the last 1~Gyr, while values above
1.5 are associated to galaxies that have finished forming their stars at least
1~Gyr ago, and are therefore evolving mostly, if not exclusively, via the
passive evolution of their stellar population.  Instead the \oii\ line is
sampling only the instantaneous star formation activity, because the ionizing
flux in H{\small II} regions is provided almost entirely by massive stars
(log(M/\msun)\,$>$\,10) with lifetimes limited to 20~Myr or less. Therefore,
the star formation time-scales probed with the \dn\ and the \oii\ equivalent
width are very different.
Despite their difference, however, there is a general agreement between them:
both indicators consistently point towards a star formation history peaked in
the past for massive galaxies, with little or no residual star formation
taking place at the epoch of the observation.  Vice-versa most of the low-mass
systems show just the opposite characteristics: almost all galaxies in this
subset have small \dn\ and a rather strong current star formation activity.

In our sample massive galaxies are the only subset of the galaxy population
that shows a significant presence of objects with large \dn, and all these
objects have also very low or null recent star formation activity.  For
low-mass galaxies both indicators suggest a general picture of extended star
formation activity. These galaxies are dominated by the young stellar
population at all epochs (although with this kind of analysis we cannot say
anything about their epoch of formation, of course).

The very good agreement observed for massive objects between the instantaneous
and the integrated star formation indicators can be understood by assuming
that the dominant mode of star formation activity for these systems after an
initial burst is a continuous one, with relatively smooth variations in time,
and with no significant contribution from a secondary strong burst. A similar
conclusion has been presented recently by \cite{noe07}, based on their
analysis of the AEGIS survey data.

We notice that a somewhat similar dependence of burst activity on galaxy
stellar mass has been observed in the SDSS data by \cite{kau03a}, although
their most recent work \citep{kau06} indicates a more important role for
galaxy surface mass density than for total mass in regulating star formation
activity.

The agreement between the two star formation indicators would be violated by
the small fraction of low-mass spectral early-type galaxies where a large \dn\
value, that points towards a predominantly old stellar population, is coupled
with a significant \oii\ emission, should this feature be indicative of
ongoing star formation activity (see Fig.\,2, panel b).
We have already listed a series of caveats on the risks of over-interpreting
the significance of such population. It is nevertheless worth mentioning that
it is possible to reproduce such a dichotomy between the two indicators when the
starburst is involving only a very small fraction of the galaxy total stellar
population (approximately less than 2\% by mass, but the exact value for this
limit depends on the amount of dimming introduced by dust extinction localised
in the star-forming regions).

The general picture which emerges from the Figs.\,\ref{fig:d4n} and
\ref{fig:count2} follows the classical concept of downsizing as firstly
suggested by \citet{cow96}.  In this framework one expects a continuous growth
in stellar mass of low-mass galaxies with time while the massive galaxies have
stopped to form stars at early epochs.  This galaxy attitude is also termed
downsizing in star formation, or in time \citep{nei06}. Most of the claims of
downsizing in the literature fall in this category \citep[\eg][]{bau05,
jun05, feu05a, feu05b, tre06}.

\begin{table*}[th!]
\begin{center}
\tabcolsep2mm
\caption{\sc {Schechter Fit Parameters}}
\smallskip        
\label{tab:median}
\bigskip
\begin{tabular}{c|clcc|clcc|c}
\hline
\hline
$z$ range     &       &  ~~~~~~~$\alpha$        & log$M^*_{star}$  & $\phi^*$                   & & ~~~~~~~$\alpha$        & log$M^*_{star}$  & $\phi^*$                   & log$M_{tr}$\\ 
              &       &                         &   ($h_{70}$~\msun)& ($10^{-3}h^3_{70}Mpc^{-3}$)& &                         &   ($h_{70}$~\msun)& ($10^{-3}h^3_{70}Mpc^{-3}$)& \\ \hline
$0.5<z<0.7$~~ &ETG~~~ &  $-0.36^{+0.19}_{-0.19} $ & $11.06^{+0.12}_{-0.10} $  & $1.50^{+0.25}_{-0.30}$ & LTG~~~ &  $-1.44^{+0.11}_{-0.09}$   & $10.96$                   & $1.05^{+0.5}_{-0.35}$  &$10.87$\\  
$0.7<z<1.0$~~ &       &  $-0.46^{+0.20}_{-0.19} $ & $11.09^{+0.12}_{-0.10} $  & $0.80^{+0.15}_{-0.15}$ &        &  $-1.34^{+0.14}_{-0.12}$   & $10.96^{+0.15}_{-0.13} $  & $1.30^{+0.6}_{-0.45}$  &$11.11$\\   
$1.0<z<1.3$~~ &       &  $-0.46                 $ & $11.20^{+0.07}_{-0.06} $  & $0.25^{+0.15}_{-0.15}$ &        &  $-1.34$                   & $11.08^{+0.07}_{-0.08} $  & $0.75^{+0.1}_{-0.10}$  &$11.42$\\   
\hline \hline\\
\end{tabular}
\begin{minipage}{17.5cm}
\normalsize{{\small NOTES:} The listed values are computed on
the early-type (ETG) and late-type (LTG) samples using a classification scheme
based on the age-estimator \dn\ (see Sect.\,\ref{sec:class}).}
\end{minipage}
\end{center}
\end{table*}

\subsection{Stellar Mass Assembly in Early- and Late-Type Galaxies} 
\label{sec:mf}

\begin{figure}[ht!]
\begin{center}
\vspace*{-0.5cm} \includegraphics[width=9.1cm,angle=0]{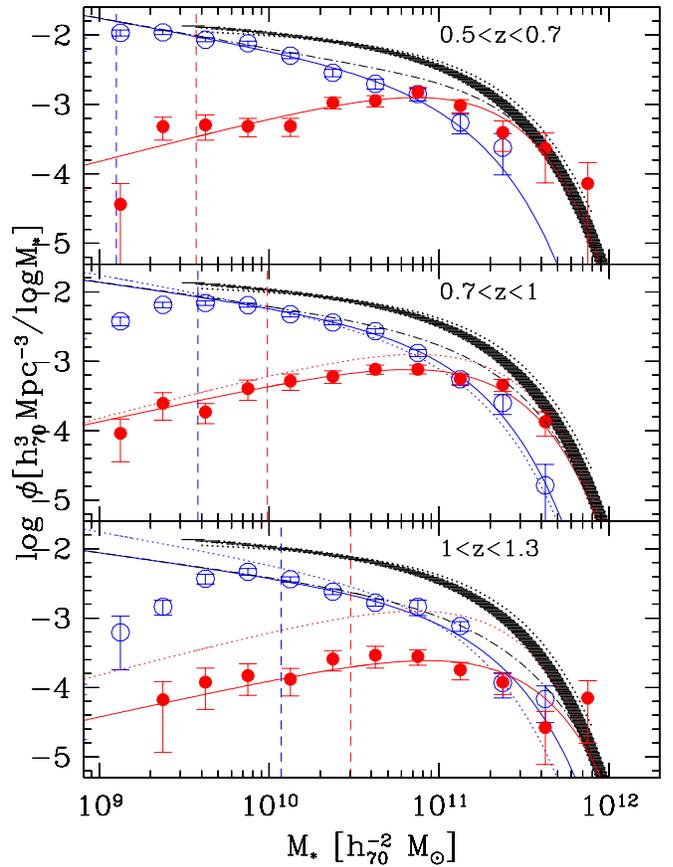}
\end{center}
\caption{Stellar mass function for the spectroscopically classified late-type
(empty circles) and early-type (filled circles) galaxies from VVDS-F02 sample.
The total mass function for each redshift bin is plotted as dashed-point line,
while in each panel we also report the total local MF range, using Cole et
al. (2001) and Bell et al. (2003) (shaded region and dotted line,
respectively).  The stellar mass functions of the two spectral types of the
lowest redshift bin are over-plotted at each panel as dotted lines. The
vertical dashed lines represent the 80\% completeness limits for the two
spectral types.}
\label{fig:mass}
\end{figure}

We show in Fig.\,\ref{fig:mass} the stellar mass function (MF) in three
redshift bins both for early- and late-type galaxies according to the spectral
classification introduced in Sect.\,\ref{sec:class}. We also show the total
mass function for each redshift bin, but we refer to P07 for a detailed
discussion on the VVDS-F02 mass function. The parameters of the Schechter fit
($\alpha,M_{star}^*,\phi^*$) for early- and late-type galaxy samples are
listed in Tab.~2 along with their uncertainties.
As the interval $0.7\le z \le1$ provides the best sampling of the overall
galaxy population we adopt the characteristic mass ($M_{star}^*$) of late-type
galaxies measured in this bin for the lower redshift one (where we do not
sample well enough the high-mass end of the population), and the slopes of both
early- and late-type MFs measured in this bin for the higher redshift one.
The vertical dashed line in each panel represents our mass limit in that
redshift bin (see Sect.\,\ref{sec:mass}). To judge the evolution in the three
redshift intervals we over-plot the Schechter fits to early- and late-type
stellar mass functions of the lowest redshift bin at each panel (red and blue,
dotted lines), and the total local MF range, using Cole et al. (2001) and Bell
et al. (2003) (shaded region and dotted line, respectively).

Comparing the total galaxy MF of Fig.\,3 to the local MF, we find a strong
evolution depending on galaxy stellar mass confirming previous results of
P07. Galaxies with stellar masses larger than log(M/\msun)\,$>$\,11 show an
evolution on the order of $30-50$\% up to {\it z}~$\sim$~0.7-1.0 and a faster
evolution above {\it z}~$\sim$~1, while the number density of less massive
galaxies decreases more continuously with redshift.

Our main focus is on the contribution of the different galaxy populations to
  the total mass function and to its evolution.
For stellar masses larger than log(M/\msun)\,$>$\,{11.4} most of the galaxies
  have an early-type classification over all the redshift range explored
  ($0.5\le z\le1.3$) and no massive late-type galaxies
  (log(M/\msun)\,$>$\,{11.5}) are observed at {\it z}~$<$~0.7 within our
  sample.
Figure\,3 also shows that at the intermediate/low-mass tail
(log(M/\msun)\,$<$\,{10}) the number density of both early- and late-type
galaxies increases with cosmic time: by a factor of $\sim$~1.6 at ${\rm
log(M)}$~$=$~10~\msun\ within the redshift range explored (and a factor of 3
compared to local MFs) for late-type galaxies, and more strongly for
early-type ones (by a factor of 5 within the redshift range explored).
We can therefore conclude that while the massive end of the mass function is
  mainly dominated by early-type galaxies up to {\it z}~$\sim$~1,
  late-type galaxies mostly contribute to the intermediate/low-mass part
  ($<{10.5}$~\msun) of the mass function at all redshifts.  Furthermore the
  abundance of massive early-type galaxies ($>$\,{11}~\msun) increases with
  cosmic time but with a declining rate at lower redshift. There is also an
  indication of a small decreasing of the massive tail of the late-type mass
  function from {\it z}~$=$~1.3 to {\it z}~$=$~0.5.

According to our results, the stellar mass function shows an increasing
(decreasing) relative contribution of massive early (late) type galaxies, and
a general increase, but at different rates and depending on the mass, of both
early and late-type galaxies with cosmic time. These results imply a
redistribution of the stellar mass amongst the spectral types with a declining
contribution to the overall mass function of massive late-type with time
associated with an increasing number density of intermediate-massive
early-type galaxies.  

The overall picture emerging from our stellar mass function agrees quite well
with previous results \citep[\eg][]{fon04,bun06}.  Generally the stellar mass
function of late-type galaxies agrees very well in all studies. As far as the
stellar mass function for the early-type galaxies is concerned the better
agreement is obtained with the study by \cite{fon04} who use a spectral
classification, instead of a rest-frame colour scheme \citep{bun06}. A
slightly worse concordance is found for the early-type galaxy densities at
{\it z}~$>$~0.7 with Bundy et al. which are larger than the one computed in
our study. Furthermore, we find a stronger evolution of early-type massive
galaxies at {\it z}~$>$~1 compared to the one measured for red galaxies in
Bundy et al.

The mass redistribution of morphological types as a function of redshift was
first noticed by \cite{be00} computing the stellar mass density in a small
sample of I-selected galaxies with spectroscopic redshifts and infrared
photometry.  Using Hubble Space Telescope morphologies, \citet{be00} observed
a decline in the mass density of irregular galaxies from {\it z}~$\simeq$~1 to
today either associated to a transformation into regular morphologies and/or
to merging mechanisms.  These results were confirmed by \citet{bun05} using a
sample of approximately two thousand morphologically-classified galaxies from
GOODS \citep{gia04}. Later on \cite{bun06} provided further support to the
stellar mass redistribution using the DEEP2 spectroscopic sample producing
type-dependent stellar mass functions based on rest-frame colours.  Further
evidence can also be found in other studies. For example, a similar modest
change in abundance of massive early-type up to {\it z}~$\sim$~1 was found in
the K20 sample which has a high redshift accuracy and a spectral
classification similar to ours \citep[][]{fon04}.  The constancy in blue
populations and the increasing trend of the red sequence classified according
to ($NUV-r$) colours have been observed in a study of stellar mass density by
Arnouts et al. (2007), and previously observed by \cite{bor06} and
\cite{mar07}.

We observe in Fig.\,\ref{fig:mass} a downward transition with time of the
threshold in the stellar mass above which the early-type galaxies become
dominant in the relative contribution to the total stellar mass function. This
characteristic mass, or transition mass ($M_{tr}$ reported in Tab.~2), was
detected in the local Universe around $3\times 10^{10}$~\msun\
\citep{kau03b,bal04}. \citet{bun06} report its evolution with cosmic time
using three different methods to define the two galaxy categories: galaxies
are partitioned according to rest-frame colours, to the star formation rate
and to morphological properties.  Our findings based on the \dn\ index are in
between those obtained dividing the galaxies by colours and by morphology in
\citet{bun06} and show a similar trend in cosmic time. A less steep evolution
is observed when considering the classification obtained by \citet{bun06} with
the SFR indicator.

\begin{figure}[hb!]
\begin{center}
\leavevmode
\includegraphics[width=9.1cm,angle=0]{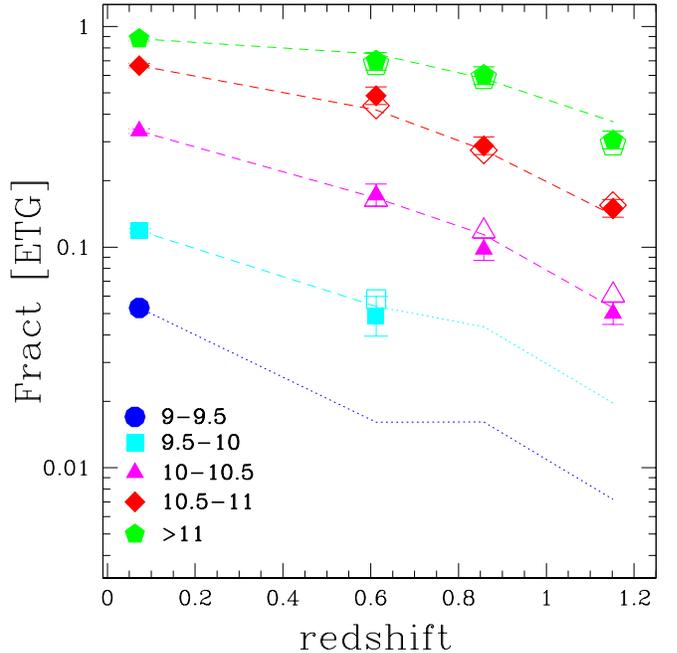}
\end{center}
\caption{The evolution of the fraction of galaxies classified
  spectroscopically as early-type in different mass intervals. The mass bins
  are coded as follows: $9.0 < {\rm log(M/M_{\odot})}\le 9.5$ (blue, circle),
  ${9.5} < {\rm log(M/M_{\odot})} \le {10}$ (cyan, square), ${10} < {\rm
  log(M/M_{\odot})} \le {10.5}$ (magenta, triangle), ${10.5}<{\rm
  log(M/M_{\odot})}<{11}$ (red, diamond), ${\rm log(M/M_{\odot})} >{11}$
  (green, pentagon). Filled symbols represent the counts of observed galaxies,
  while the open symbols represent the counts derived from 1/V$_{max}$
  formalism. The dashed lines are the Schechter fitted values to the stellar
  mass function (dotted in the case of mass completeness for early-type
  galaxies below 50\%.)  The points at {\it z}~$\sim$~0 are our SDSS reference
  points.}
\label{fig:count1}
\end{figure}

To better show the differential evolution of galaxies and the general
redistribution in stellar mass among galaxies we plot in
Fig.\,\ref{fig:count1} the evolution of the early-type galaxy fraction over
the entire (early- and late-type) population for different stellar mass bins
as a function of redshift. Both observed counts and Schechter fitted values
are used in the computation and plotted with different symbols. The error-bars
include both the uncertainties in the spectral measurements and the effect of
changing the classification due to the \dn\ errors. As in Fig.\,\ref{fig:d4n}
we report the reference points at {\it z}~$\sim$~0 from a local galaxy sample
(SDSS DR4).

Figure\,4 clearly shows how the evolution of the relative abundances of early-
and late-type galaxies strongly depends on galaxy stellar mass. The relative
fraction of the most massive galaxies (log(M)~$>$~{11}~\msun) significantly
changes with cosmic time, between {\it z}~$=$~1.2 and {\it z}~$=$~0.85 due to
both the decrease of massive late-type and the increase of massive early-type
galaxies, followed by a flattening below {\it z}~$=$~0.85, when this second
population becomes dominant in this mass regime.  A similar flattening is
present at {\it z}~$<$~0.6 for galaxies with log(M)~$>$~{10.5-11}~\msun. A more
continuous evolutionary trend is suggested for the relative fraction within
the less massive population (${<10.5}$~\msun): both late- and early-type
densities increase but the first category of galaxies is always dominant at
this stellar mass regime (Fig.\,\ref{fig:mass}).

As these measured ratios could be affected both by different mass completeness
limits for early and late-type galaxies (as the VVDS is a magnitude limited
survey) and by any incompleteness in the global spectroscopic sample, we derive
the relative fraction of early-type galaxies also integrating the best fit
Schechter function. The agreement between the observed ratios and the ones
derived by the Schechter integration shows that our results are robust.

To summarise, our results suggest a galaxy mass assembly history with a mild
evolution with redshift in number density (P07) for very massive galaxies
($>$~{11.4}~\msun), which are mainly constituted by early-type galaxies up to
the redshift sampled ({\it z}~$<$~1.3). 
Since the relative fraction of massive early- and late-type galaxies evolves
with cosmic time we witness a spectral transformation of late-type systems
into old massive galaxies at lower redshift.  In Sect.\,\ref{sec:efficency} we
analyse the efficiency of this mass assembly process driven by the star
formation rate observed in galaxies.

\subsection{Influence of cosmic variance}
\label{sec:cv}

An intrinsic source of uncertainty in any estimate of number densities from a
finite-size survey is represented by the variance induced by density
fluctuations on scales comparable to, or larger than, that of the sampled
volume. This is what is commonly termed ``cosmic variance''. It inversely
depends on the sample volume and directly on the two-point correlation
function of the class of galaxies under analysis.  This cosmic variance will
introduce a systematic effect on the overall amplitude of our mass function.
The significance of this effect needs to be evaluated.  This can be obtained
in our case with fairly good accuracy for the three redshift bins of VVDS-F02
used here, since we already know the clustering properties of both the early-
and late-type galaxy population over the whole range covered by the survey
\citep{men06}.

Following \citet{pee80} and \citet{som04}, under the assumption of a
power-law correlation function $\xi(r) = (r/r_0)^{-\gamma}$ the {\it
  relative} variance of galaxy counts over a given volume with radius
$R$ can be expressed as
\begin{equation}
  \sigma^2_R = J_2\, (R/r_0)^{-\gamma} \,\,\,    .
\label{eq:variance}
\end{equation}
\noindent Here $J_2$ is a classical integral over the two-point correlation
function as defined in \citet{pee80}, which in the power-law
approximation can be expressed as  
\begin{equation}
J_2 = 72/[(3-\gamma)(4-\gamma)(6-\gamma)2^\gamma] \,\,\, .
\label{eq:J2}
\end{equation}
In the case of a non-spherical volume $V$, we can nevertheless define an
equivalent radius as $R=(3V/4\pi)^{1/3}$.  For our three samples, this
corresponds to $R=34$, 45 and 49 h$^{-1}$ Mpc, respectively.  From
\citet{men06}, we know that, to sufficient approximation, the correlation
length $r_0$ and slope $\gamma$ of $\xi(r)$ do not evolve significantly, being
$r_0\simeq 4$ $(2.5)$ h$^{-1}$ Mpc and $\gamma=1.8$ $(1.5)$ for early- (late-)
type galaxies respectively, over the full redshift range under investigation.
Thus, we have all the ingredients to use eqs.~\ref{eq:variance} and
\ref{eq:J2} to estimate the expected {\it rms} relative error on the measured
densities, $\sigma_{counts}=\sqrt{\sigma^2_R}$, for each of the three samples.
The results are reported in Table~\ref{tab:variance}.  The listed values
indicate that the systematic error on our estimates of the mass due to cosmic
variance is expected to be between 13 and 20\%.  This is fully consistent with
the result we have obtained by splitting the sample into two parts and
recomputing the mass functions separately for the two subsamples. The
agreement between these two independent estimates allows us to conclude that
our conclusions are robust with respect to cosmic variance.

\begin{table}
\tabcolsep4.5mm
  \caption{Relative {\it rms} error on galaxy number
    densities expected from cosmic variance in the three redshift samples}
  \label{tab:variance}
  \centering
  \begin{tabular}{l | c c c }
    \hline \hline
         &\multicolumn{3}{c}{$\sigma_{counts}$}\\
    \hline
    Type & $0.5<z<0.7$ & $0.7<z<1$ & $1<z<1.3$ \\
    \hline
    Early & 0.20 & 0.15 & 0.14 \\
    Late & 0.17 & 0.14 & 0.13 \\
    \hline
  \end{tabular}
\end{table}

\begin{figure*}
\begin{center}
\includegraphics[width=18cm,angle=0]{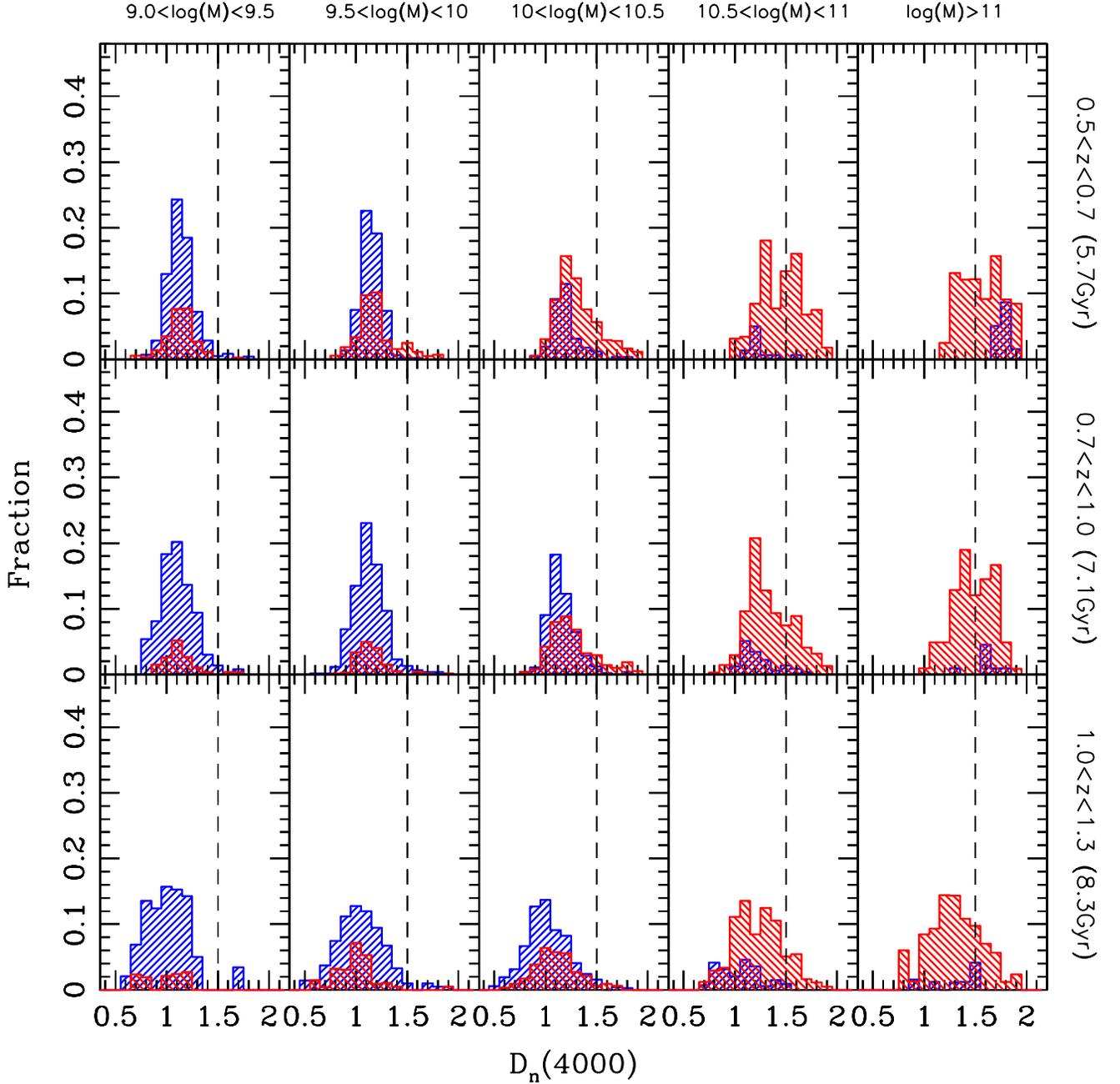}
\end{center}
\caption{Histograms showing the fraction of VVDS galaxies as a function of
\dn\ for five different ranges of stellar mass at three redshift epochs
($0.5-0.7, 0.7-1.0, 1.0-1.3$).  The corresponding look-back time is also
indicated.  We define $\Delta t$ for each galaxy at redshift {\it z} as the
time interval between the age of the Universe at the observed redshift {\it z}
and the time corresponding to the lower limit of the redshift bin in which the
galaxy falls.  Left-leaning diagonal (blue) histograms represent those
galaxies that assemble within a time $\Delta t$ a stellar mass enough to move
to a higher mass interval and with right-leaning diagonals (red) those staying
in the mass interval assigned at the time of the observations.  The assumption
is that these galaxies will sustain a constant observed SFR over the time
$\Delta t$, which turns to be a reasonable assumption as our $\Delta t$ is
never larger than 1.3~Gyr.}
\label{fig:doubling}
\end{figure*}

\subsection{Efficiency in the Stellar Mass Assembly Process}
\label{sec:efficency}

In the case of dry merging - merging of quiescent galaxies without gas (star
  formation) involved - the age of the galaxy (\ie the time elapsed since it
  assembled most of its current stellar mass) may not coincide with the age of
  its underlying stellar population. As a consequence the processes of star
  formation and mass assembly remain distinct: it is possible that stars which
  formed long ago were assembled only recently in a newly created galaxy
  \citep{ren07, cim06, bun06}.

Within such a scenario one considers two separate downsizing signatures, \ie
downsizing in star formation (the transfer of star formation activity to lower
mass galaxies) and downsizing in mass assembly (irrespectively by type
classification, massive galaxies are fully assembled at early times, with less
massive galaxies assembling later).  If one or more mechanisms are responsible
for quenching star formation activity earlier in massive galaxies, and if
dry mergers represent only a minor contribution to the galaxy mass assembly,
then the two concepts of downsizing express one single idea.  In the previous
sections we have found that the distribution of the stellar ages and stellar mass
can be interpreted in the framework of a top-down picture of the stellar mass
assembly history of galaxies. However, we made no hypothesis on the mechanism
or mechanisms that lead the stellar mass to aggregate, leaving open the
possibility of mass accretion by both merging and star formation activity.

Can the galaxy star formation rates and their efficiencies alone justify the
observed mass assembly history without invoking any (dry) merger mechanism?
To investigate this issue we define for each galaxy at redshift {\it z} the
time $\Delta t$ as the time interval between the age of the Universe at the
observed redshift {\it z} and the time corresponding to the lower limit of the
redshift bin in which the galaxy falls.  For three redshift bins in Fig.\,5 we
plot with left-leaning diagonals (blue-coded) the \dn\ distribution of
galaxies which assemble enough stellar mass to move to a higher mass interval
only on the basis of their estimated SFRs within a time $\Delta t$. The
assumption is that galaxies will sustain a SFR, equal to the observed one,
over the time $\Delta t$. This appears to be a reasonable assumption as our
$\Delta t$ is never larger than 1.3~Gyr. The right-leaning diagonals
(red-coded) represent the galaxies which, given their observed SFR, would
remain over the time $\Delta t$ in the mass interval assigned at the time of
the observations.

Figure~\ref{fig:doubling} shows that the lower the stellar mass the larger is
the fraction of galaxies which sustain a significant stellar mass growth, \ie
not only forming stars as a local and episodic event but also effectively
assembling mass.  Once a galaxy reaches a certain stellar mass, it terminates
to grow efficiently.  The physical reason could be the natural exhaustion of
the gas reservoir contained in massive galaxies which did consume it
efficiently in the first phase of their life. On the contrary, low-mass
galaxies, having sustained a lower rate of star formation activity over the
time still have a certain amount of fueling.

The fraction of galaxies that increase significantly their stellar mass
(blue-coded population) is larger than the fraction of galaxies that is not
effectively assembling mass (red-coded) at lower stellar masses with
time. The evolution of the characteristic stellar mass at which blue- and
red-coded populations are equally contributing to the total population
parallels that of the quenching mass, defined as the mass at which no active
systems are observed \citep[see for details][]{bun06}.

For each stellar mass bin Fig.\,5 shows a decline in the mass assembly
  efficiency from high to low redshifts. Similarly for each redshift bin there
  is a decline in the mass assembly efficiency moving from low to high stellar
  masses.  Interestingly, the few massive galaxies which can still potentially
  increase their mass are classified as late-types from \dn.

These results and those obtained in previous sections support the hypothesis
of apparent no evolution in the blue sequence proposed by Faber et al. (2007)
and the mild evolution of the early-type class of galaxies \citep[][ Arnouts
et al. 2007]{zuc06,bro07} which are largely inefficient at assembling mass at
all epochs (at least since {\it z}~$\sim$~1.3), and are progressively decreasing their
efficiency as cosmic time progresses.  In particular, on the basis of the mass
migration from blue to red objects, Arnouts et al. (2007) propose a similar
time-scale for the period required for blue galaxies to quench their star
formation activity and move to the red sequence, and the birth rate time of
blue galaxies which are continuously refilled by new stars.

If we consider the galaxies with stellar mass M$_i$
(log(M/\msun)$_{i={1,2,3,4,5}} = 9-9.5, 9.5-10, 10-10.5, 10.5-11, >11$)
observed at redshift $z_j$ ($z_{j={1,2,3}} = 0.5-0.7, 0.7-1, 1-1.3$), their
progenitors are the sum of two galaxy contributions if no merging activity is
advocated.  The first progenitor class is constituted by galaxies with mass
M$_{i-1}$ at $z_{j-1}$ that are assembling enough stellar mass to migrate in
the higher mass interval within a time $\Delta t$, where $\Delta t$ was
previously defined (e.g. blue-coded galaxies at {\it z}~$=$~0.7-1 and
log(M)~$=$~{10.5-11}~\msun).
The second contributors are galaxies with stellar mass included in the same
mass interval (M$_{i}$) at the epoch $z_{j-1}$ that in a time $\Delta t$ are
not growing sufficient mass to flow in the subsequent stellar mass interval
(e.g. red-coded galaxies at {\it z}~$=$~0.5-0.7 and log(M)~$>$~{11}~\msun).

We compute within our mass completeness limit the number (per unit of
co-moving volume) of galaxies observed at redshift $z_j$ of stellar mass M$_i$
and that of progenitors at redshift $z_{j-1}$ as described above.  We obtain
that the number of progenitors at $z_{j-1}$ defined as above can account, on
average, for (80$\pm10$)\% of the galaxies at $z_j$ of stellar mass M$_i$, and
therefore that no significant contribution to the mass assembly process is
required from merging activity.  The agreement between number of progenitors
and number of observed galaxies improves moving to higher masses, reaching
nearly 100\% for log(M)~$>$~{10.5}~\msun.  This result shows that star
formation activity alone can explain the observed evolution of stellar mass
function up to {\it z}~$\sim$~1.
A similar conclusion has been presented recently by \cite{bun07} for
spheroidal galaxies comparing the dynamical and stellar mass in the GOODS
fields.  The match between the growth rate in the abundance of spheroidals
with that predicted by the assembly of dark matter halos enables Bundy et
al. to state that merging is not the primary mechanism to form spheroids and
additional mechanisms as morphological transformations are required to drive
the observed evolution.
Further convincing evidence that a significant number of massive blue galaxies
must have quenched their star formation and moved to the red sequence without
invoking any merging mechanism is provided by \eg \citet{sca07} and
\citet{bel07}. This picture satisfies both the luminosity and mass function
constraints.

The scenario described here, if correct, implies that all recent observations
on galaxy evolution can be summarised with one single concept of downsizing,
as downsizing in star formation and downsizing in mass assembly go in
parallel.  Still, a more detailed and quantitative analysis of the mass
function and star formation rate evolution are needed before a more accurate
evaluation of the role played by mergers in the build up of the galaxy mass
can be obtained. It is in fact well known that mergers do take place and
recently individual cases of dry mergers have also been listed by \eg
\citet{bel06} and \citet{tra05}. We cannot therefore exclude the possibility
that between two redshift bins a small fraction of galaxies moves from mass
bin M$_i$ to a higher mass bin via merger.

A possible objection to the scenario in which merging would not be important
is the large fraction of galaxies up to {\it z}~$=$~1 possessing spatial and
kinematic asymmetries \citep[e.g.][]{flo06, pue07}. These
lopsided galaxies are very often associated to merging of galaxies. However
there exist various ranges of mass merging fractions and several mechanisms
(slow accretion, gravitational interaction, etc.) that can justify the
observed asymmetric properties and are not probed by the current surveys.  It
could very well be that major mergers -- that are those that we are
constraining with our analysis -- are not needed to explain the observed
lopsidedness frequency.  These surely are issues to investigate with future
(gas multi-phase) surveys.

\section{Summary}

We have investigated the relationship between galaxy stellar age and stellar
mass as a function of redshift in a mass-complete spectroscopic sample
selected from the VIMOS VLT Deep Survey (VVDS). Up to {\it z}~$\sim$~1.3 we
confirmed the presence of a relationship mass-stellar ages that parallels the
one observed in the local Universe (see Kauffman et al. 2003b). In all
redshift bins explored, low-mass galaxies (log(M)~$<$~{10}~\msun) are dominated
by a young stellar population, as witnessed by the low \dn\ values. For higher
mass galaxies (log(M)~$>$~{10}~\msun), the percentage of galaxies dominated by
an old stellar population is much higher and grows regularly with cosmic
time. This process is more efficient the higher the galaxy stellar mass
considered. This result supports the so-called 'assembly downsizing': the
stellar population in massive galaxies formed at earlier times than the one in
low-mass galaxies.

We then explored SFR evolution as a function of stellar mass. The percentage
of quiescent galaxies, as witnessed by their low EW\oii\ values, increases
when moving to lower redshifts and higher masses. This trend is clearly
visible both for galaxies with a young stellar population and for galaxies
with an old stellar population. The emerging picture is one where low-mass
galaxies (log(M)~$<$~{10}~\msun) are subject to bursts of star formation
activity up to recent times, while the dominant mode of star formation for
massive galaxies (log(M)~$>$~{10}~\msun) is a smooth one, with no significant
contributions from secondary strong bursts. Therefore not only does the bulk
of the stellar population in massive galaxies form at earlier times but the
location where the mass also gets assembled more efficiently at each redshift
moves to lower mass galaxies with cosmic time.

We have also shown a mild total evolution with redshift of mass assembly
history, in particular for the very massive galaxies (log(M)~$>$~{11.4}~\msun)
which are mainly early-type objects, up to the highest sample redshift ({\it
z}~$\sim$~1.3). Since the relative fraction of massive
(log(M)~$>$~{10.5}~\msun) early- and late-type galaxies evolves with cosmic
time, we are witnessing a spectral transformation of late-type systems into old
massive galaxies at lower redshift.

Finally, when we consider the joint distribution of stellar mass and star
formation activity to quantify the efficiency with which galaxies assemble
their stellar mass, we obtain a scenario where it is possible to account for
the number of passively evolving galaxies up to {\it z}~$\sim$~1 without
invoking any (dry) merging mechanism.  Our observations agree well with a
scenario where the stellar population in massive galaxies formed earlier than
the one in low-mass galaxies, and massive galaxies themselves were the first
to be assembled.

\section{Acknowledgments}

This research has been developed within the framework of the VVDS consortium.
This work has been partially supported by the CNRS-INSU and its Programme
National de Cosmologie (France), and by Italian Ministry (MIUR) grants
COFIN2000 (MM02037133) and COFIN2003 (num.~2003020150). The VLT-VIMOS
observations have been carried out on guaranteed time (GTO) allocated by the
European Southern Observatory (ESO) to the VIRMOS consortium, under a
contractual agreement between the Centre National de la Recherche Scientifique
of France, heading a consortium of French and Italian institutes, and ESO, to
design, manufacture and test the VIMOS instrument. DV acknowledges the support
through a Marie Curie ERG, funded by the European Commission under contract
No. MERG-CT-2005-021704. 
Based on observations obtained with MegaPrime/MegaCam, a joint  
project of CFHT and CEA/DAPNIA, at the Canada-France-Hawaii Telescope  
(CFHT) which is operated by the National Research Council (NRC) of  
Canada, the Institut National des Science de l'Univers of the Centre  
National de la Recherche Scientifique (CNRS) of France, and the  
University of Hawaii. This work is based in part on data products  
produced at TERAPIX and the Canadian Astronomy Data Centre as part of  
the Canada-France-Hawaii Telescope Legacy Survey, a collaborative  
project of NRC and CNRS.

\bibliographystyle{aa} \bibliography{biblio}

\end{document}